\documentclass{aipproc}

\newcommand{\GeV}{\,{\rm GeV}}
\newcommand{\I}{{\em I}}
\begin{document}

\title{Looking for Instanton-induced Processes at HERA Using a Multivariate Technique\\Based on Range Searching}
\author{T. Carli\affiliationmark{1}
  and B. Koblitz\thanks{Talk
    given at the ACAT 2000 conference.} \affiliationmark{ 1}\affiliationmark{,2}}
\affiliation{\affiliationmark{1} Universit\"at Hamburg
  and \affiliationmark{2}MPI f\"ur Physik, F\"ohringer Ring 6,
  D-80805 M\"unchen}

\begin{abstract}
  We present a method to discriminate instanton-induced processes from
  standard DIS background based on Range Searching. This method offers
  fast and automatic scanning of a large number of variables for a
  combination of variables giving high signal to background ratio and
  the smallest theoretical and experimental uncertainties.
\end{abstract}

\maketitle

\section{Instanton-induced processes}
Instantons \cite{Belavin} are a fundamental non-perturbative aspect of
QCD, inducing hard processes that are absent in perturbation theory.
The expected cross section as calculated in
``instanton-perturbation-theory'' is sufficiently large \cite{
  MRS, RS2, RS3} to make an experimental discovery
possible \cite{RS1,CGRS}. For a more detailed introduction to instantons
(\I) see e.g. \cite{RS4}.

We study the prospect of a search for \I-induced events modelled by
the Monte Carlo Generator QCDINS \cite{RS5} which generates \I-induced
events in deep-inelastic $ep$-scattering where a quark emerging from a
$q\bar q$-splitting of the exchanged photon fuses with a gluon emitted
from the proton.  In the \I-induced process $q\bar q$-pairs of each of
the three light quark flavours and on average 2-3 gluons are
produced.  In the hadronic CMS they form a band (of about two units in
pseudo-rapidity) of particles with high transverse energy which are
homogeneously distributed in azimuth. Since in every event a pair of
strange quarks is produced, in this band an increased number of kaons
compared to standard DIS events is expected. Finally, the quark out of
the split photon not participating in the instanton subprocess
forms a hard jet.

The predicted cross section $\sigma_{\rm
  HERA}^{(I)}=29.2^{+9.9}_{-8.1}\,\rm pb$ \cite{RS3,RS4} in a kinematic
region where ``instanton-perturbation-theory'' ($x_B>10^{-3},\,
0.1<y<0.9,\, Q^2>113\GeV^2$) is applicable, is two orders of magnitude
smaller than the DIS cross section $\sigma_{\rm DIS}\approx 3000\,\rm
pb$. Therefore the highest possible signal to background ratio has to
be achieved by exploiting observables characterising \I-induced
processes. To find these observables a large number of promising event
variables have to be investigated and the sensitivity to systematic
details in the modelling of the hadronic final state has to be tested.
This requires a sophisticated and fast discrimination method to find
the appropriate combination of event variables.

\begin{figure}[t]
\includegraphics[width=15cm]{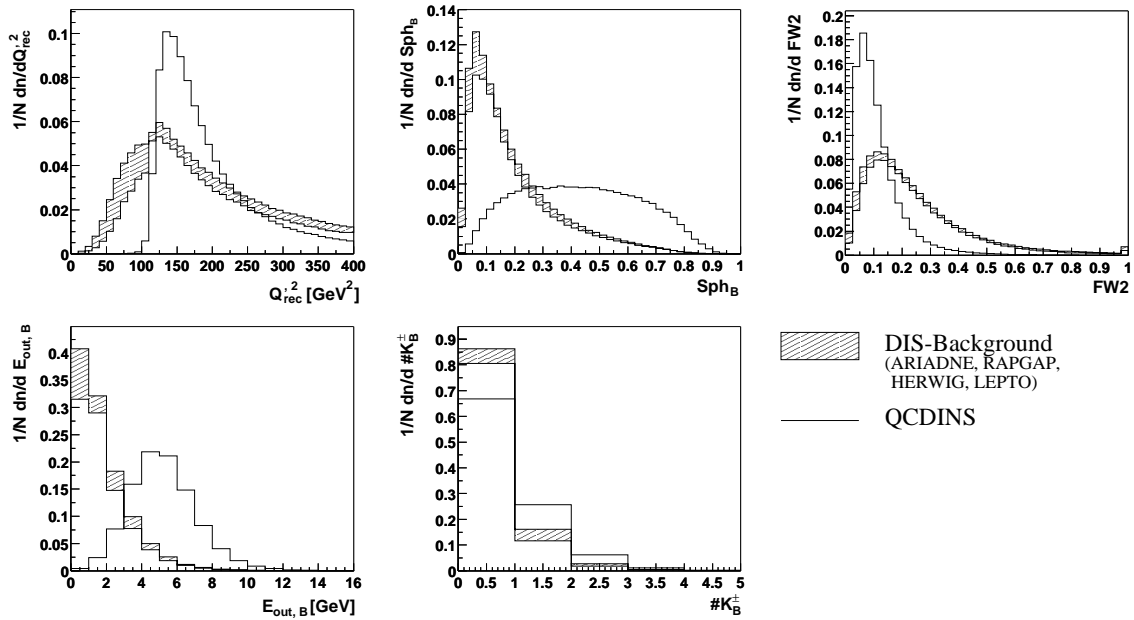}
\caption{The characteristic event variables providing good
  instanton separation with small systematic uncertainties. Shown is
  the reconstructed virtuality of the quark entering the \I-subprocess
  $Q_{\rm rec}'^2$, the sphericity of the particles in the \I-Band in
  their rest system, the second Fox-Wolfram moment of these particles
  and the event shape variable $E_{\rm out, B}$ which is the
  projection of the particle transverse energy onto the axis, that
  makes this quantity maximal (see \cite{CGRS}). Finally the number of
  charged kaons in the \I-Band is shown.  }
\label{fig:a}
\end{figure}

\section{Range Searching}
Events can be classified as signal or background by estimating the
probability density $\rho$ of both these classes at the point of the event in
the event-variable phase space, employing a Monte Carlo (MC) generator to
sample the densities. In the case of neural networks (NN) this is done by
fitting the probability-density with the adjusted weights of the
neurons. To circumvent this time consuming procedure the density at
each point can be directly estimated by counting the number of
background and signal events in a surrounding box $V$. Given the ratio
$$\ell:=\frac{\rho(I)}{\rho(DIS)}=\frac{\#I(V)}{\#DIS(V)}$$
the probability of an event to be
a signal event is $D=\ell/(1+\ell)$. Compared to NN's this method also
has the advantage of not extrapolating into phase space regions where
there are no sample events available. Thus signals from data
events outside the region covered by the MC simulation can be
avoided. This is not the case for NN's which extrapolate
into regions where there is no test data. Counting the number of
events in the vicinity of a certain point is a problem known as Range
Searching.

Range Searching algorithms have been developed which allow a search
time $\sim \log(n)$, where $n$ is the number of points that have to be
searched \cite{Sedg}. We employ an algorithm \cite{BK} suitable
especially for a large number of events and dimensions (i.e.
observables).  The MC events are successively filled into the nodes of
two binary trees, one for the signal events and one for the background
events, where the criterion by which the decision is taken to descend
to the left or right of a node is given by the value of one of the
event variables.  While descending the tree this variable cycles
through all the ones considered.  After filling, the position of every
event in the tree is given by its coordinates in the event variable
space.  Classification of an event is done by searching in the trees
for all background and signal events in the box $V$. This is done in
the same manner as filling the tree.

\begin{figure}[t]
\includegraphics[width=16cm]{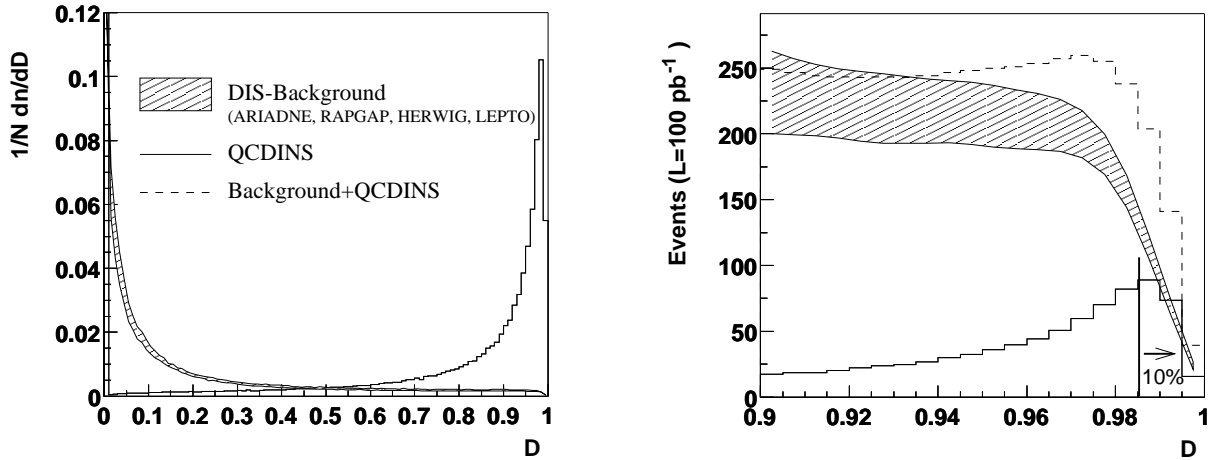}
%\parbox{.7\textwidth}{\mbox{}\vspace*{3cm}}
\caption{To the left, the shape normalised discriminant $D$ for the
  instanton events using QCDINS and for standard DIS events using four
  MC simulators is shown. The second plot shows a zoom into the rightmost
  part and is now normalised to a luminosity of $L=100\,{\rm
    pb^{-1}}$. At $\epsilon(I)=10\%$ 178 \I-induced events are
  expected.}
\label{fig:b}
\end{figure}

\section{Results}
Starting with 35 variables based on the hadronic final state the best
12 were chosen by calculating the discriminant with all 2-combinations
(pairs) of the initial variables and taking those variables which
provide a high separation power $S=\epsilon(I)/\epsilon({DIS})$
demanding an efficiency for instantons of $\epsilon(I)=10\%$. The
number of considered variables is further reduced by calculating all
5-combinations and selecting those with highest separation power and a
small systematic variation of the background.  The systematic
uncertainty was obtained by using four standard DIS-MC simulators \cite{MCs}
which were tuned to data on representative hadronic final state
quantities, in the range $Q^2>100\GeV^2$ at HERA \cite{NBrook}. The
variables forming the best combination is shown in Figure~\ref{fig:a}.
The separation power for $\epsilon({\rm I})=10\%$ is $S=126$. In
Figure~\ref{fig:b} the shape normalised discriminant $D$ is shown for
the \I-induced and the background events, as well as the distribution
for $D>0.9$ normalised to a luminosity of $100\,{\rm pb^{-1}}$ which
is comparable to that already collected by each of the HERA experiments
H1 and ZEUS. An event sample can be isolated where half of the events
are instantons while the \I-efficiency is still 10\%.

\par
\begin{figure}
\includegraphics[width=8cm]{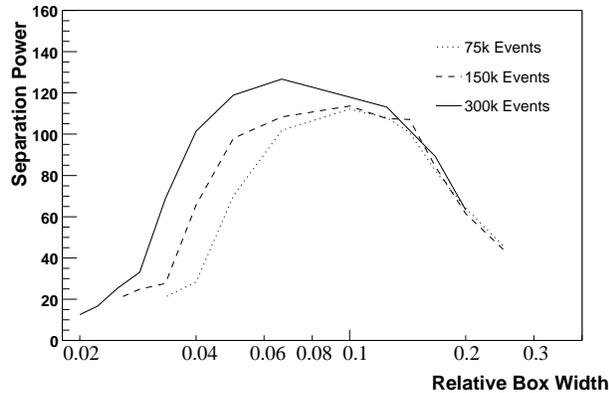}
%\parbox{.7\textwidth}{\mbox{}\vspace*{3cm}}
\caption{The separation power S at $\epsilon(I)=10\%$ for different box
  sizes and different numbers of events in the search trees.}
\label{fig:c}
\end{figure}

For a search method to be reliable and easy to apply it is important
to have as few free parameters as possible. In the case of Range
Searching these are the number of events in the search trees, the size
of the neighbourhood $V$ and the minimum number of events in this
neighbourhood to classify an event. To reduce the number of parameters
for the box size the ratios of the box edge lengths were fixed by
defining a box which contains most of the events and letting $V$ be a
scaled version of this large box. The projections onto these box edges
are shown in Figure~\ref{fig:a}. The variation of the result depending
on the size of $V$ is shown in Figure~\ref{fig:c}. Clearly the
separation increases for smaller boxes with the number of events that
populate the search trees, while for larger boxes this difference
vanishes. The plateau is increasing in width with the number of tree
events and reaches nearly an order of magnitude in size, thus allowing
to use only an approximate size parameter and reducing the need for
fine tuning, if enough MC statistics is available.

In addition a comparison with a single hidden layer feed forward NN
was done. The network performing best had 100 hidden nodes and was
trained with the same input data. It reached a separation of $S=116$
at an \I-efficiency of 10\%, being slightly worse than the Range
Search method. Training the net was rather time consuming
\footnote{$4\,{\rm h}$ compared to $20\,{\rm min}$ for the Range
  Search method on a Linux PC} and a lot of human intervention had to
be done to adjust the input scales and training parameters.

\section{Conclusions}
The multivariate discrimination method based on Range Searching
performs at least as good as a NN's when applied to the search for
instantons at HERA. It is much less time consuming and can be easily
used to automatically scan a large number of appropriate variables.
The short processing time allows extensive searches for the best
discriminating variables taking systematic effects into account. In a
region where \I-perturbation theory can be safely employed this novel
discrimination method results in an 50\% \I-enriched data sample while
the \I-efficiency is still 10\%.

\section{Acknowledgments}
We would like to thank A. Ringwald and F. Schrempp for their fruitful
and delighting collaboration and for many stimulating discussions.

%%%%%%%%%%%%%%%%%%%%%%%%%%%%%%%%%%%%%%%%
% look into test.bbl for examples to manually generate references
% the stuff below interfaces BibTeX
%%%%%%%%%%%%%%%%%%%%%%%%%%%%%%%%%%%%%%%%
\nocite{*}
\bibliographystyle{aipproc}
\bibliography{test}
\end{document}